# Motor vehicles accidents and teenage drivers: A statistical analysis of their age and injuries

Debo Brata Paul Argha [1], * and Md Javed Imtiaze Khan [2]

[1] Ingram School of Engineering, Texas State University, San Marcos, Texas, 78666, USA; arghapaul7148@gmail.com

[2] Ingram School of Engineering, Texas State University, San Marcos, TX 78666, USA; m_k352@txstate.edu

* Correspondence: arghapaul7148@gmail.com

**Abstract**

Motorcycle accidents are a prevalent problem in Texas, resulting in hundreds of injuries and deaths each year. Motorcycles provide the driver with little physical protection during accidents compared to cars and other vehicles, so when there is a collision involving a motorcycle, the motorcyclist is likely to be injured. While there are numerous reasons for motorcycle accidents, most are caused by negligence and could have been avoided. Because of the increasing popularity of motorcycles and scooters in Texas, coupled with an increase in the number of motorcycle accidents, the Texas Department of Transportation (TXDOT) has "amped" its efforts to improve motorcycle safety. From the data, it has been visible that teenage drivers are the most vulnerable to motorcycle accidents. In this report, we have tried to find out the probability of young driver and passenger motorcyclist's injury based on different conditions and to predict the rate of changing injury to this group in the upcoming years.

### Introduction

Road accidents are one of the top incidents of fatality of Texas citizens. Road crashes can be classified into several groups depending on the vehicle involved and its occupant capacity. Among these motorcyclist fatalities and injuries are responsible for a larger portion of all losses. As a source of data, we will go through TxDOT's (Texas Department of Transportation) database of Crash Data analysis and Statistics. All the data are categorized into six main types based on the injury. They are Fatalities, Suspected – Serious Injuries, Non-Incapacitating Injuries, Possible Injuries, Not Injured and Unknown Injuries. Among all the types of fatalities the biggest point is where all losses are maximum. Additionally, the database of Crash Data analysis and Statistics is further classified based on the injury that happened to the driver and passengers too. Both the driver and the passenger category can be further classified based on the condition of the helmet use. These are: 'Worn', Not worn', and 'Unknown'. So, if the mathematical formula of combination is applied then there is a total of 54 different conditions can be considered to analyze an accident. Furthermore, the set of data is vast when the age group is taken into consideration too. Age starting from 0 to 99 years is listed accordingly mention the count of injuries. For instance, a branch view of the conditions is represented in figure 1. Here only the data of 2019 is shown only for the fatality incident.

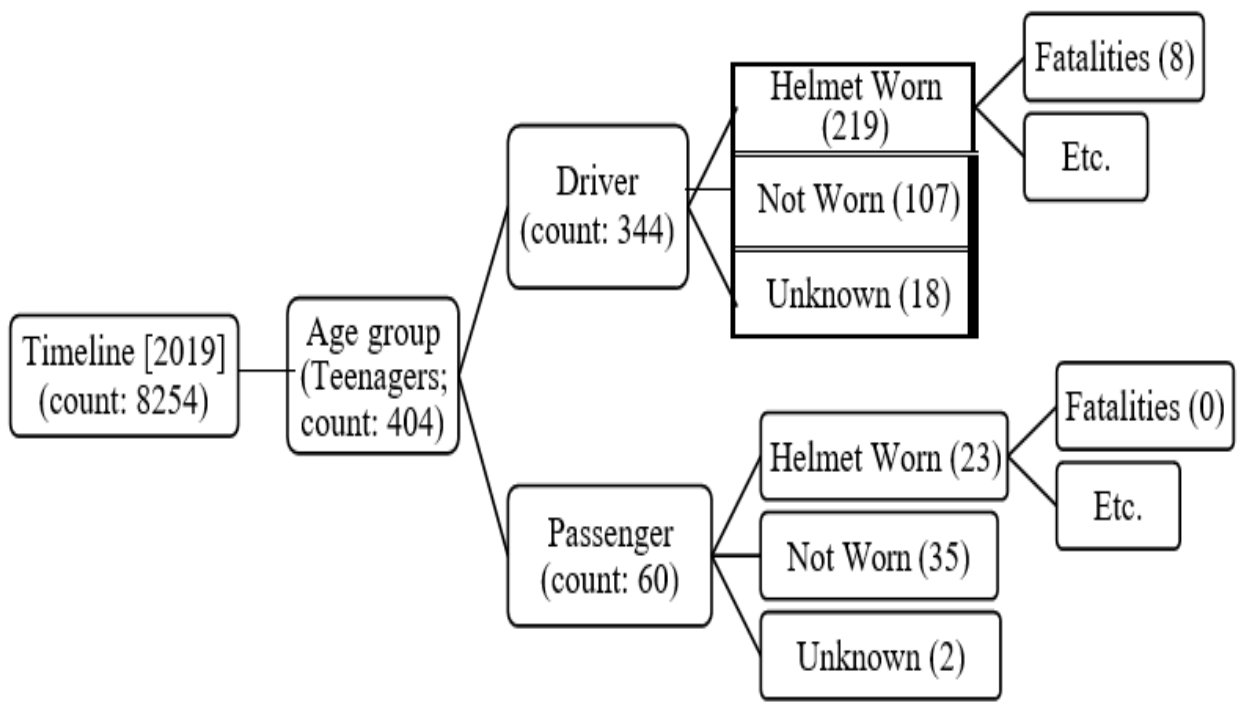

Figure 1: a branch view of the accidental incident in 2019 for teenager fatality.

To understand the trend of injury, meaningful mathematical representations are must from these huge data. In addition, these will give the insight to predict future fatalities trend of the Texas state too. this mathematical model will be useful to understand and predict incidents too of other states if the necessary information is available. Motorcycle accident starting from 2015 to 2019 is available open to access on the source, but more can be accessed with approval from the proper authority. The summary of the objectives of this project is listed below:

- ❖ We figured out the teenager rider's probability of fatalities and injuries depending on different conditions.

- ❖ To predict future motorcyclist fatalities for the next three (03) years.
- ❖ To implement different probability formulas for determining the result on mutually dependent and independent situations.

**Methodology & DataAnalysis**

At first, the five years fatality data have been plotted to the Minitab Software to find out the normal statistical parameter. From Minitab, it has been also checked whether this data follow any distribution pattern (Normal Distribution) (Figure 2). For driver fatality rate, a one-dimensional linear equation has been used to forecast the fatality in coming years. The single exponential method has been used to forecast where smoothing constant α (alpha) = 0.56 (Figure 3). For this calculation, a 95% confidence level has been considered. Some procedure has been followed to forecast the passenger fatality rate for the coming years by developing one-dimensional linear equation and single exponential method. In this study, a time series plot has been used for understanding age-wise fatality comparison. This total age limit has been divided into four categories (Figure 3).

Normality Test of Total Number of Driver Fatality

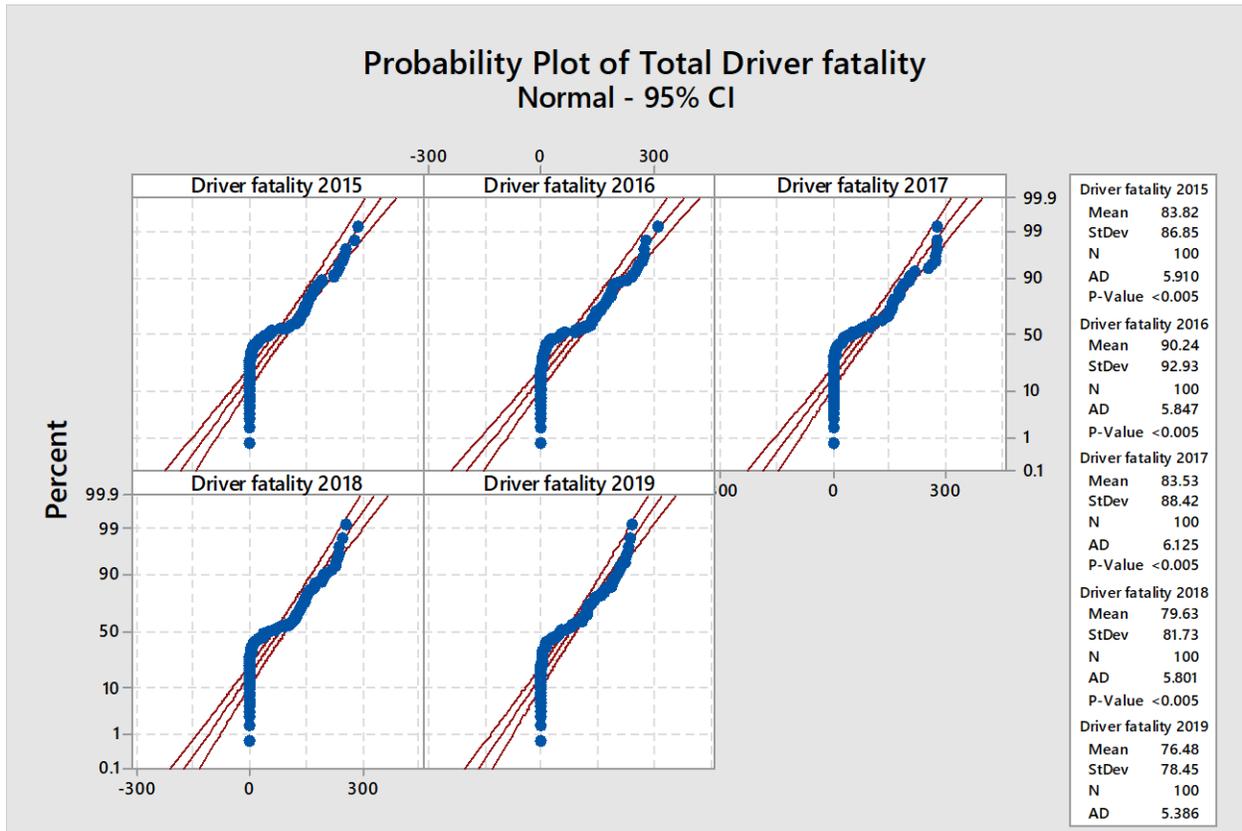

Figure 2: Probability plot of total driver fatality.

From the graph, it is shown that the driver fatality of 2015, 2016, 2017, 2018, 2019 do not follow the normal distribution since P-value for every case is less than .005. For five different years, it is also seen that the Mean is highest for the year 2016 and lowest for the year 2019. For standard deviation, it is highest in the year 2016 and lowest in the year 2019. For the year 2017, the Anderson Darling value is highest and for the year 2019, it is the lowest among these five years of data.

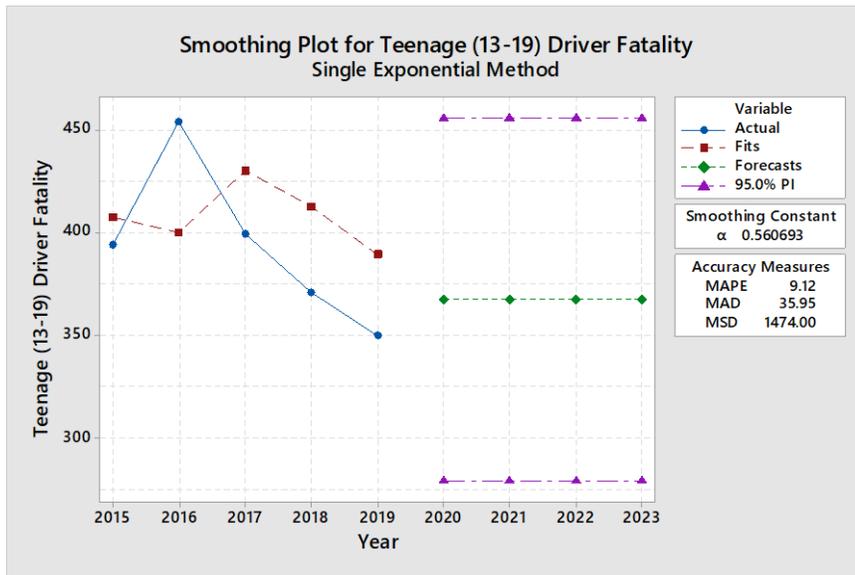

(a)

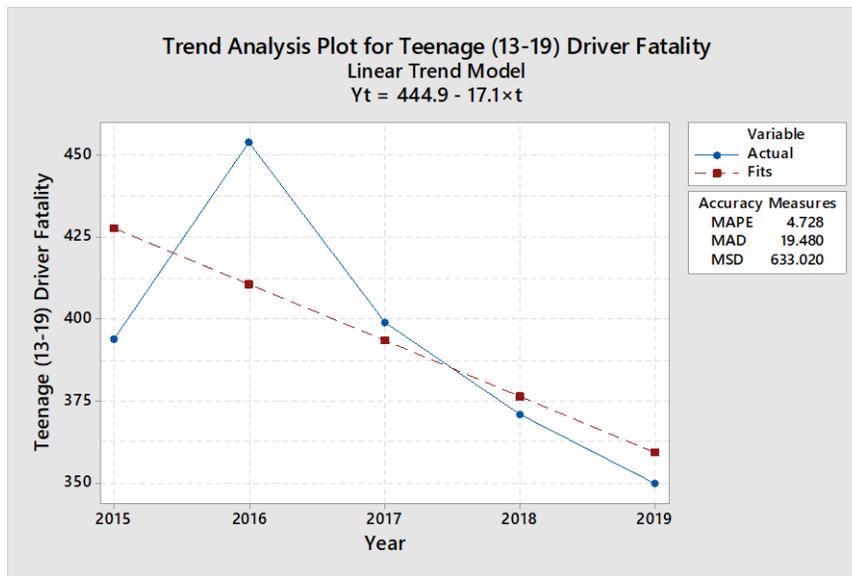

(b)

Figure 3: (a) Smoothing plot for teenage driver fatality (b) Trend analysis plot for teenage driver fatality

From the above graph, we have developed a trend and a linear equation of teenage driver fatality.

We also used the Smoothing plot for Teenage (13-19) Driver fatality by Single Exponential method. Here, From the trend analysis graph we see, MAPE (Mean absolute percentage error) is 4.728 and MAD (Median absolute deviation) is 19.480 and MSD (Mean squared deviation) is 633.020. From the Smoothing plot graph we see, MAPE (Mean absolute percentage error) is 9.12 and MAD (Median absolute deviation) is 35.95 and MSD (Mean squared deviation) is 1474.00.

**Data Teenage (13-19) Driver Fatality Forecast**

Length 5

Smoothing Constant α 0.560693

Accuracy Measures

| MAPE | MAD | MSD |
|------|------|------|
| 9.12 | 35.95 | 1474.00 |

Teenage (13-19) Driver

| Time | Fatality | Smooth | Predict | Error |
|------|----------|--------|---------|--------|
| 2015 | 394 | 399.874 | 407.370 | -13.3700 |
| 2016 | 454 | 430.222 | 399.874 | 54.1264 |
| 2017 | 399 | 412.716 | 430.222 | -31.2219 |
| 2018 | 371 | 389.326 | 412.716 | -41.7160 |
| 2019 | 350 | 367.276 | 389.326 | -39.3261 |

**Forecasts**

| Period | Forecast | Lower | Upper |
|--------|----------|-------|-------|
| 2020 | 367.276 | 279.195 | 455.357 |
| 2021 | 367.276 | 279.195 | 455.357 |
| 2022 | 367.276 | 279.195 | 455.357 |
| 2023 | 367.276 | 279.195 | 455.357 |

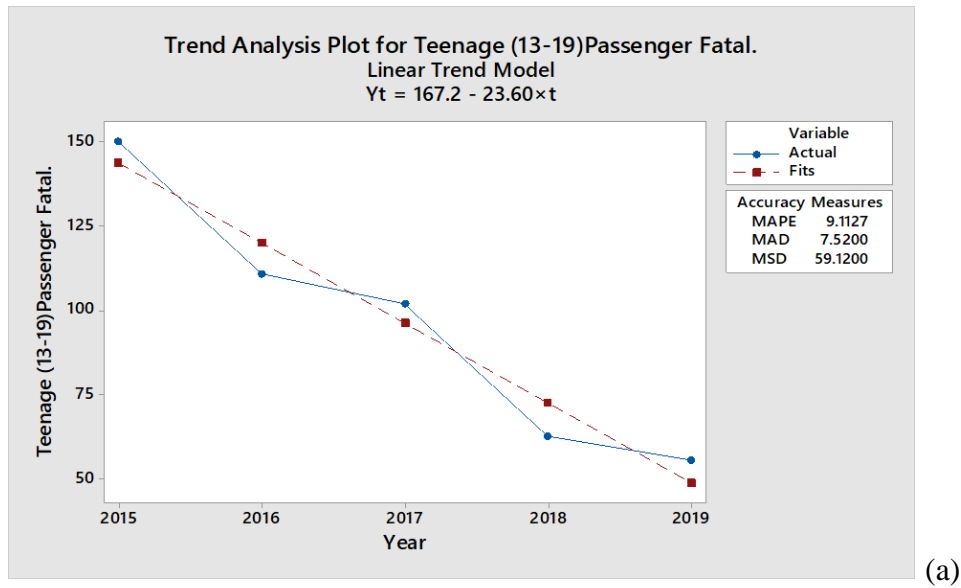
(a)

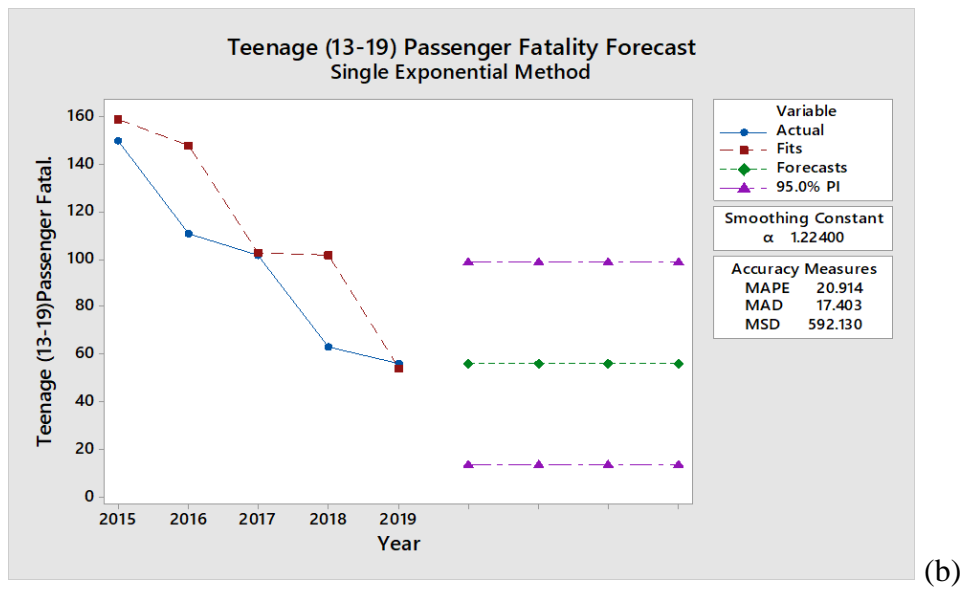
(b)

Figure 4: (a) Trend analysis plot for teenage driver fatality (b) Teenage Passenger fatality forecast.

In Figure 4, we have developed a trend analysis for teenage (13-19) passenger fatality and a linear equation. We have forecasted by using the single exponential method. Here, MAPE (Mean absolute percentage error) is 9.1127 and MAD (Median absolute deviation) is 7.5200 and MSD (Mean squared deviation) is 59.1200. From the Smoothing plot graph we see, MAPE (Mean absolute percentage error) is 20.914 and MAD (Median absolute deviation) is 17.403 and MSD (Mean squared deviation) is 592.130.

**Data   Teenage (13-19) Passenger Fatality forecast**

Length 5

Smoothing Constant α 1.22400

Accuracy Measures

| MAPE | MAD | MSD |
|------|-----|-----|
| 20.914 | 17.403 | 592.130 |

**Teenage (13-19) Passenger fatality**

| Time | Fatal. | Smooth | Predict | Error |
|------|--------|--------|---------|-------|
| 2015 | 150 | 148.046 | 158.723 | -8.7227 |
| 2016 | 111 | 102.702 | 148.046 | -37.0461 |
| 2017 | 102 | 101.843 | 102.702 | -0.7017 |
| 2018 | 63 | 54.299 | 101.843 | -38.8428 |
| 2019 | 56 | 56.381 | 54.299 | 1.7007 |

**Forecasts**

| Period | Forecast Lower Upper | Period | Forecast Lower Upper |
|--------|---------------------|--------|---------------------|
| 2020 | 56.3810 | 13.7448 | 99.0171 |
| 2021 | 56.3810 | 13.7448 | 99.0171 |
| 2022 | 56.3810 | 13.7448 | 99.0171 |
| 2023 | 56.3810 | 13.7448 | 99.0171 |

As we only consider 5 years of data, the single exponential method unable to develop a trend since data is very limited and as a result it gives the same value again and again. But we can guess that the fatality rate will be around the above mention value in the coming three or four years.

In this figure 5, we have developed a trend analysis (Linear Trend Model) and developed a fatality forecast single exponential method.

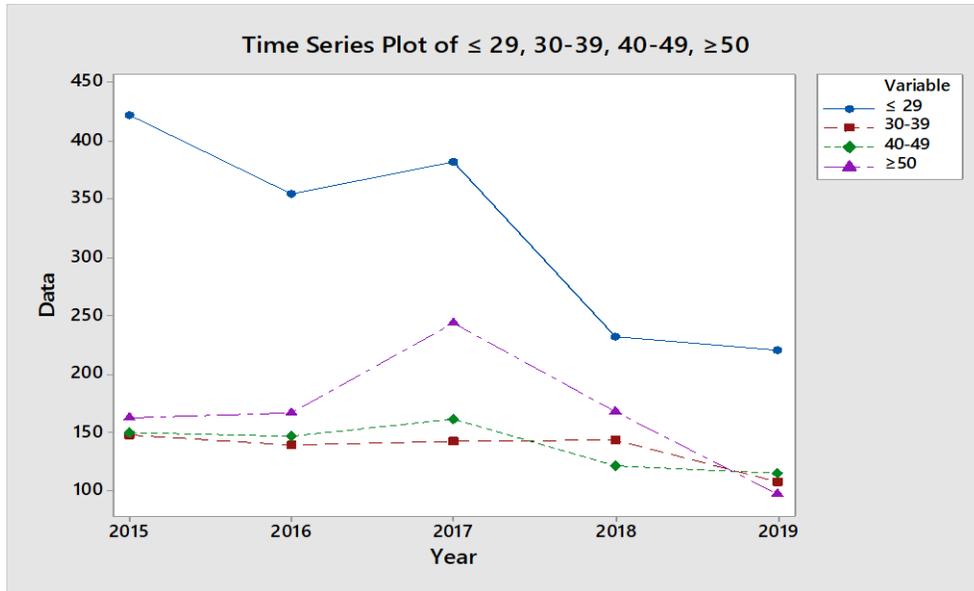

Figure: 5

In figure 5, we have shown the year-wise fatality. 1st case age limit is less than 29, 2nd case between 30-39, 3rd case 40-49, 4th case 40-49. From figure 5, it is visible that passenger and driver below 29 years are very susceptible to injury. For the age between 30-39, the fatality is almost constant. From the above figure 5, it is visible that the fatality is downward for all age groups which is one of the key findings of this study. A sample calculation of the probability of teenage driver being selected for fatality case is shown here (statistical data 2019 is used for this calculation):

Teenage in accident, $\quad\quad\quad\quad$ P(T) $\;$ = 0.049
Among them, Driver $\quad\quad\quad\quad$ P(D) $\;$ = 0.042
$\quad\quad\quad\quad$ Passenger $\quad\quad\quad\quad$ P(P) $\;$ = 0.007
$\quad\quad\quad\quad$ Driver Fatalities $\quad\;$ P($D_f$) = 0.027
$\quad\quad\quad\quad$ Passenger Fatalities $\;$ P($P_f$) = 0.003

According to the Law of total probability, if an incident is randomly selected, the probability of being a teenage fatality is

$$= P(D_f|D).P(D) + P(P_f|P).P(P)$$
$$= 0.27*0.42 + 0.003*0.007$$
$$= 0.113$$

**Discussion & Conclusion**

To conclude the study, it can be said that the fatality rate for teenage drivers is high and even the fatality rate is high for the age below 29 years both for passengers and drivers. The fatality rate for passengers is somewhat constant in recent years. Moreover, in the case of Passengers and drivers over 50, the fatality rate is almost the same, and we can conclude that they are less vulnerable to the motorcycle accident. All the objectives of this study are tried to fulfill. Due to time limitation, sample calculation is made using selective data from the whole, but a more in-depth picture can be made if each condition is taken into consideration.